\documentclass[preprint,12pt]{elsarticle}
\usepackage{xcolor}
\usepackage{textcomp}

\usepackage[export]{adjustbox}[2011/08/13]

\usepackage{caption}
\usepackage{subcaption}

\usepackage{multicol}
\usepackage{color, colortbl}
\usepackage{multirow}
\definecolor{LightBlue}{rgb}{0.9,0.9,0.9}

\usepackage{graphicx}
\usepackage{amssymb}

\usepackage[symbol]{footmisc}

\usepackage{amsmath}

\usepackage{footnote}
\makesavenoteenv{tabular}
\makesavenoteenv{table}

\usepackage{siunitx}

\usepackage{hyperref}

\usepackage{cleveref}

\usepackage{lineno}

\biboptions{sort&compress}

\newcounter{bla}

\journal{Computer Physics Communications} 

\begin{document}

\begin{frontmatter}


\title{ADBSat: Methodology of a novel panel method tool for aerodynamic analysis of satellites}



\author[a]{Luciana A. Sinpetru\footnote{Corresponding author. \\ \textit{E-mail address:} luciana.sinpetru@manchester.ac.uk}}
\author[a]{Nicholas H. Crisp}
\author[b]{David Mostaza-Prieto}
\author[a]{Sabrina Livadiotti}
\author[a]{Peter C.E. Roberts}

\address[a]{The University of Manchester, Oxford Road, Manchester, M13 9PL, United Kingdom}
\address[b]{Hispasat, Calle de Anabel Segura 11, 28108 Alcobendas, Madrid, Spain}

\begin{abstract}

ADBSat is a novel software that determines the aerodynamic properties of any body in free-molecular flow. Its main advantage is the fast approximation of the aerodynamics of spacecraft in the lower end of the low-Earth orbit altitude range. It is a novel implementation of a panel method, where the body is represented as a set of fundamental elements and the sum of their individual aerodynamic properties makes up the properties of the whole. ADBSat's approach treats the shape as a set of flat triangular plates. These are read from a CAD geometry file in the Wavefront format, which can be created with most common CAD programs. A choice of gas-surface interaction models is available to represent the physics of free-molecular flow under different conditions. Its modular design means that other models can be easily and quickly implemented. It also benefits from a new shading algorithm for fast determination of elemental flow exposure. 
An example case is presented to show the capability and functionality of the program. 
\end{abstract}


\end{frontmatter}

\newpage

{\bf PROGRAM SUMMARY}

\begin{small}
\noindent
{\em Program Title:} ADBSat                                          \\
{\em CPC Library link to program files:} (to be added by Technical Editor) \\
{\em Code Ocean capsule:} (to be added by Technical Editor)\\
{\em Licensing provisions:} GPLv3\\
{\em Programming language:} MATLAB \\
{\em Nature of problem:} Quickly and accurately determining the aerodynamics of satellites in free-molecular flow.\\ 
{\em Solution method:} A new implementation of the panel method has been devised. The satellite shape is passed to the program as a CAD model, comprised of a set of flat triangular plates. ADBSat then calculates the aerodynamic characteristics of each element using an appropriate mathematical model, and sums the contributions for the overall properties of the body. A novel shading algorithm identifies and removes the panels which do not contribute to the calculations due to being protected from the flow by other body features. ADBSat also has the capability to account for different materials within the shape. \\ 
{\em Additional comments including restrictions and unusual features:} MATLAB's Aero-space Toolbox is required for the determination of environmental parameters, unless otherwise provided by the user. As the program takes as an input a completed model of the spacecraft, the user is responsible for all mesh quality checks. The methods employed therein are only valid for strict free-molecular flow, which the user must ensure. The accuracy of the method decreases for surfaces with high concavity or multiple particle impingement.\\ 
   \\

\end{small}


\section{Introduction}
\label{S:intro}

The panel method technique is an established method of calculating the aerodynamic properties of satellite geometries in rarefied flows, such as those in the low-earth orbit (LEO) regime \cite{FreeMat,orion,apollo,moe2004simultaneous}. In essence, it involves reducing a model of the spacecraft into a number of simple segments, and calculating the individual aerodynamic contribution of each segment. The contributions are then summed to give the aggregate aerodynamic properties of the body.

Panel methods are just one of the options available for aerodynamic analysis. They are an analytical method, applying closed-form equations to determine the drag contribution of each building block of the body. While relatively simple to implement, they have historically been disadvantaged by the difficulty in determining panels that are shielded from the flow \cite{FreeMat} and handling concave shapes that promote multiple particle reflections. Options for aerodynamic analysis that can handle these shortcomings are numerical methods, such as Direct Simulation Monte Carlo (DSMC) and Test Particle Monte Carlo (TPMC) \cite{DragModelling}, which rely on modelling the movements of particles in a simulation domain. The trade-off for any such method is the high computational complexity. They have become increasingly popular in recent years with the improvement in power of computational facilities and the advent of assistive schemes, such as interpolation and response surface modelling, that decrease the number of simulations necessary \cite{MEHTA2014_RSM}. However, even with these improvements to numerical methods, panel methods are still the quickest way to gain an understanding of the aerodynamics of a realistic satellite body.

ADBSat is a novel implementation of the established panel method which aims to overcome some of the aforementioned shortcomings. Its aim is to output a useful approximation of the aerodynamic properties of complex spacecraft within the limitations of panel methods. Additionally, it is simple and easy to implement, while further reducing computational time compared to previous applications. One common feature between this and previous applications \cite{FreeMat} is that it is primarily built within the MATLAB\textsuperscript{\tiny\textregistered} environment, making use of the programming language's matrix-based methods to reduce computational time and load. However, it is distinct in that it does not limit the user to a single gas-surface interaction model (GSIM) to describe the physics of spacecraft flight in the rarefied atmosphere. Instead, six of the most widely used GSIMs for space applications have currently been implemented, allowing the user to choose among these to best represent their specific scenario. The modular structure of the program supports the straightforward addition of others. Thus, the program has been purposefully built to be adaptable to further developments in our knowledge of the physics of free-molecular flow (FMF).

Additionally, unlike previous implementations, ADBSat does not have an in-built method for building or otherwise describing satellite shapes. Instead, it takes as an input a model created using any computer-aided design (CAD) program, output in the common Wavefront file format, which is processed into a set of triangular plates for analysis. This makes the geometry easily transferable across different modelling and simulation programs, and thus renders any obtained results easily verifiable. This will undoubtedly prove useful in the context of a wider mission design framework, in which multiple aspects of the design must be integrated in a concordant way.

The long-standing problem of efficiently calculating panel exposure to the flow has also been tackled in a novel way. Due to the hyperthermal flow assumption, which assumes that the bulk gas velocity is much greater than the individual thermal velocities of the particles, the panels that are shielded from the flow by upstream components are assumed to have zero aerodynamic pressure. In other words, they do not contribute to the aerodynamic properties of the spacecraft. This is in line with previous similar implementations \cite{FreeMat, FreeMac}. However, a new shading algorithm is used to identify these panels \cite{MostazaThesis}, based on a 2-dimensional projection of the triangular elements that the program uses to represent the geometry. There is no need for bounding boxes or subdividing the geometry multiple times, as has been previously done. 
While still being the most computationally expensive algorithm employed by the program, the novel algorithm significantly speeds up processing time. Thorough validation of the shading algorithm, including a discussion of its limitations, is presented in an accompanying paper \cite{myValidationPaper}.

\begin{figure}[h!]
     \centering
     \begin{subfigure}{0.49\textwidth}
         \centering
         \includegraphics[width=\textwidth]{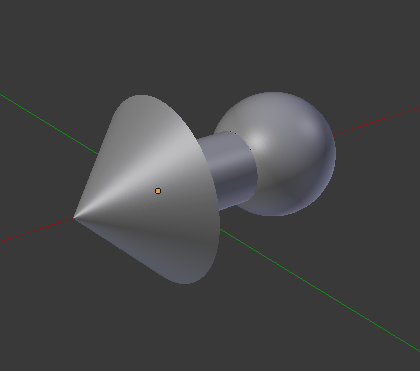}
         \caption{Arrow shape made using Blender.}
         \label{fig:arrow_blender}
     \end{subfigure}
     \hfill
     \begin{subfigure}{0.49\textwidth}
         \centering
         \includegraphics[width=\textwidth]{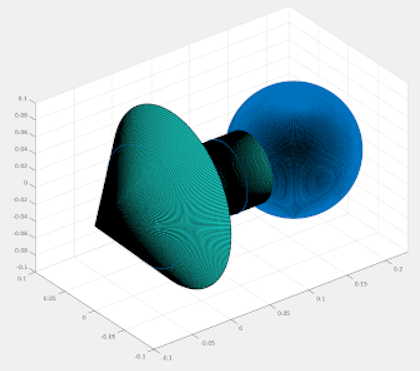}
         \caption{ADBSat import of Blender mesh geometry.}
         \label{fig:arrow_adbsat}
     \end{subfigure}
     \hfill
     
     \vspace{0.5cm}
     
     \begin{subfigure}{0.60\textwidth}
         \centering
         \includegraphics[width=\textwidth]{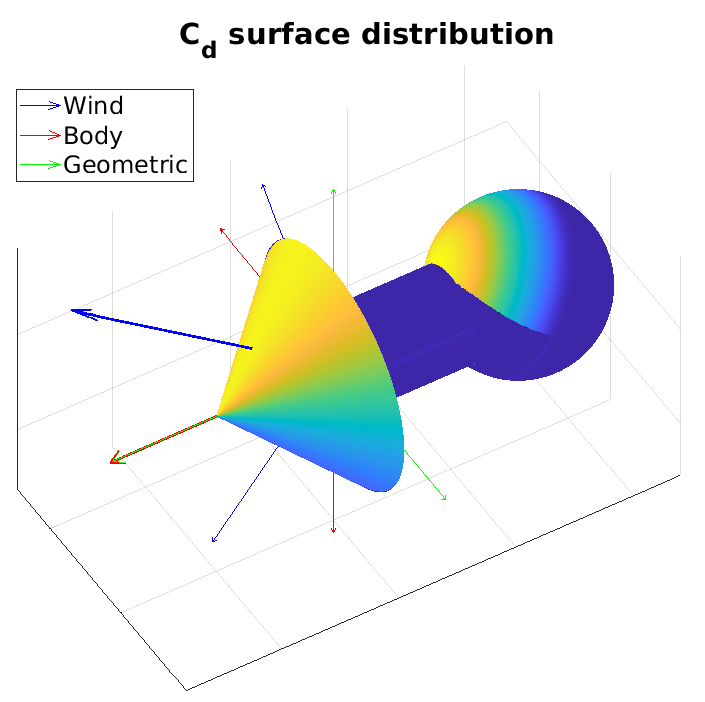}
         \caption{Analysis at an angled flow demonstrates the shading algorithm. The indigo panels do not contribute to the aerodynamic coefficients.}
         \label{fig:arrow_shading}
     \end{subfigure}
        \caption{Analysis of an example geometry showing the progression from CAD modelling software, to ADBSat import, to aerodynamic analysis involving the shading algorithm.}
        \label{fig:arrow}
\end{figure}

\Cref{fig:arrow} shows the steps of importing an example arrow geometry and analysing its aerodynamic properties using the shading determination algorithm. This shape was first designed using the Blender CAD software \cite{blender} as shown in \cref{fig:arrow_blender}. It was then imported into ADBSat through the included import function, the results of which can be seen in \cref{fig:arrow_adbsat}. Furthermore, on examining the aerodynamics of the shape at an angled flow, the effects of the shading algorithm can be seen in \cref{fig:arrow_shading}. In this image, the indigo panels are omitted from the final aerodynamic calculation.

Furthermore, an optional solar coefficient calculation for the body is available to the user. This addition to ADBSat has not been present in previous implementations. The same model geometry is utilised as for the aerodynamic analysis. From a systems engineering perspective, this consistency between geometric models across multiple analyses is desirable for accuracy, rather than using different representations in different software. The solar coefficient calculation can be used to determine the solar radiation pressure accelerations generated on the centre of mass of the spacecraft, which is invaluable for orbit determination. This is currently an optional extra which has yet to be fully tested and validated.

\begin{figure}[t]
     \centering
     \begin{subfigure}[b]{0.49\textwidth}
         \centering
         \includegraphics[width=\textwidth]{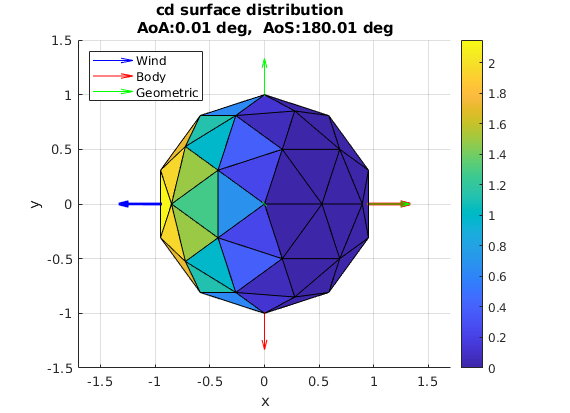}
         \caption{Drag coefficient, $C_d$}
         \label{fig:ADBSat_out_cd}
     \end{subfigure}
     \hfill
     \begin{subfigure}[b]{0.49\textwidth}
         \centering
         \includegraphics[width=\textwidth]{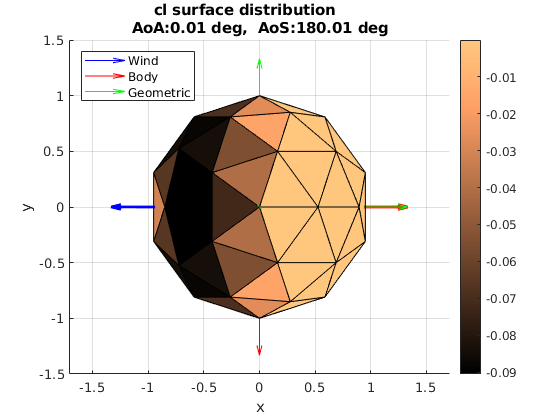}
         \caption{Lift coeffiient, $C_l$}
         \label{fig:ADBSat_out_cl}
     \end{subfigure}
     \hfill
     \begin{subfigure}[b]{0.49\textwidth}
         \centering
         \includegraphics[width=\textwidth]{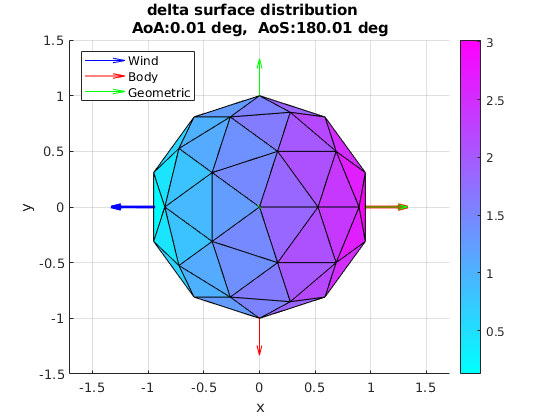}
         \caption{Angle between the plate and the flow, $\delta$ (rad)}
         \label{fig:ADBSat_out_delta}
     \end{subfigure}
        \caption{Example graphical outputs of ADBSat for a quasi-spherical polyhedron.}
        \label{fig:ADBSat_out}
\end{figure}

The outputs, which can include solar and aerodynamic coefficients, can be either for a single orientation or a database of values at various incidence angles. They are saved in a MATLAB workspace variable file (with a ".mat" extension). This file also contains relevant parameters that have been used in the calculations, such as the angle of attack ($\alpha$), angle of sideslip ($\beta$), projected area (${A_{proj}}$), reference area ($A_{ref}$) and length ($L_{ref}$). These files can either be loaded into MATLAB and viewed, or accessed directly through a MATLAB program to use the values. There is also an optional graphical output that can display the distribution of plate angle, or any of the coefficients of drag, lift, pressure, and shear stress, across the object. An example of this graphical output is shown in \cref{fig:ADBSat_out}, for a quasi-spherical polyhedron.

This work details the implementation and theory behind ADBSat, with a focus on novel contributions. The methodology is thoroughly explained. Potential uses alongside an example case are also discussed. Extensive testing and validation are addressed in an accompanying paper \cite{myValidationPaper}.

\section{Background}

The bottom end of the LEO regime, also known as very low Earth orbit (VLEO), has become an increasingly attractive orbital regime for extended satellite missions in recent years. Usually described as \SIrange{100}{450}{\kilo\metre}, this range of orbital altitudes provides varied benefits for space platforms, in particular those with a focus on Earth observation \cite{RobertsEtAl}. However, it also suffers from a significant drawback: aerodynamic drag caused by interaction of the spacecraft with the residual atmosphere. The characterisation of drag at these altitudes has been attempted by various authors throughout the years \cite{sentman, SchaafCham,Cook,storchHyp, MoeMoe1995, MoeMoe1998, MoeMoe2005, AccommodationCoeffPardini, paddlewheel, DragModelling}. Yet, it is only recently that we have begun to more thoroughly understand this unique challenge on low-altitude spacecraft flight.

Satellite flight in VLEO occurs mostly in FMF, as characterised by the Knudsen number, $K_n$, given in \cref{eq:Kn}. In this equation, $\lambda$ is the mean free path of the particles in the residual atmosphere, and $L$ is the characteristic length scale of the object.

\begin{equation}
    K_n = \frac{\lambda}{L}
    \label{eq:Kn}
\end{equation}

The Knudsen number defines the flow regime as follows \cite{DragModelling}:

\begin{itemize}
    \item $K_n \leq 0.1$: continuum flow
    \item $0.1 < K_n < 10$: transitional flow
    \item $K_n \geq 10$: FMF
\end{itemize}

 
 Another criterion for FMF can be defined through the number density of the atmosphere, $n_0$. Although less commonly used than the Knudsen number, $n_0 \le 10^{16}$ can signify FMF \cite{ComparingDragCoeffs_GSIs}. While the exact value of $K_n$ and $n_0$ are dependent on the instantaneous atmospheric conditions, the criteria for FMF are generally satisfied in VLEO \cite{DragModelling, ComparingDragCoeffs_GSIs}.

Due to the small number of intermolecular collisions, the aerodynamic interaction is instead dominated by the collision of the gas particles with the surface. The physical theory behind these collisions is captured in the GSIMs. A range of  GSIMs have been developed that make fundamentally different assumptions about the particle-surface interactions and the re-emission characteristics. The accuracy of these assumptions is related to characteristics such as the environmental conditions and the materials present \cite{sabrina,DragModelling}. Most models rely on an energy accommodation coefficient $\alpha_E$, though momentum accommodation coefficients are also used, and some models use a mixture of both for shear and normal contributions. These relate the GSIM to the material by quantifying the extent of energy or momentum transfer from a gas molecule to the surface \citep{DragModelling}. $\alpha_E$ is given by \cref{eq:accomCoeff}, and is always in the range $0 \leq \alpha_E \leq 1$. In this equation, $E_i$ is the energy of the incoming particles, $E_r$ is the energy of the reflected particles, and $E_w$ is the energy that the reflected particles would have had, had they been completely accommodated to the temperature of the surface. The momentum accommodation coefficient, $\sigma$, is calculated in a similar manner.

\begin{equation}
    \alpha_E = \frac{E_i - E_r}{E_i - E_w}
    \label{eq:accomCoeff}
\end{equation}

Due to the well-documented adsorption of atomic oxygen onto the materials of satellites orbiting in VLEO \citep{MoeMoe2005}, accommodation is highest where atomic oxygen concentration is highest. As the atmosphere becomes more rarefied with height, $\alpha_E$ decreases. Thus, $\alpha_E$ for current materials used in spacecraft production will depend mostly on altitude. Atmospheric conditions such as solar activity also influence its value. It is left to the user to select suitable values of $\alpha_E$ to provide as inputs to the program. The estimation of these coefficients often presents problems in terms of accuracy. Estimation methods include extrapolating from values obtained from laboratory measurements in different conditions than VLEO \cite{labExperiments1,labExperiments2,labExperiments3}, or from empirical experiments with limited observability \cite{Moe1966,MoeMoe2011,MoeBowmanSpheres,moe2004simultaneous,MoeMoe1998}.  \citet{AccommodationCoeffModel} have also devised a Langmuir isotherm model that can be used to calculate the accommodation coefficient for a particular set of atmospheric parameters. This model has not been validated for all altitudes and values, but it is useful in cases where its output is $\alpha_E \geq 0.85$. Widely used and accepted values for LEO range from 0.6 to 1.0 \cite{MoeMoe1998,MoeMoe2005,accomCoeff_accepted}. Research is ongoing into materials that are resistant to adsorption of atomic oxygen and promote specular reflection. However, at the time of writing, there is no experimental data available for these novel materials.

A GSIM provides equations that can be used to analytically calculate the drag coefficient of a body, $C_d$. The drag force, $F_d$, can then be calculated using \cref{eq:dragForce} by using the atmospheric mass density $\rho$ and the speed of the object relative to the surrounding gas $u$.

\begin{equation}
F_d = \frac{1}{2} \rho u^2 A_{ref} C_d
\label{eq:dragForce}
\end{equation}
%
However, an accurate estimate of $C_d$ can often be difficult to obtain, due to the significant uncertainty surrounding atmospheric conditions. 
The flow parameters, $\rho$ and $u$, may vary unpredictably across orbits. In particular, $u$ is normally assumed to be the speed of the satellite's orbit around the Earth, with the assumption of a co-rotating atmosphere. The unpredictable nature of thermospheric winds can decrease the accuracy of this assumption \cite{ThermosphereUncertainty}. There is also a significant uncertainty associated with atmospheric density modelling that must be considered \cite{atmosphericUncertainty_vallado}. Furthermore, the somewhat arbitrary choice of $A_{ref}$ can further impact accuracy. Historically, this has been defined as the cross-sectional area of the object perpendicular to the flow direction (i.e. the projected area) \cite{XsectionalArea}. For complex spacecraft shapes in a variable flow, this may not be accurately determinable at all times. As the selection of $A_{ref}$ and $C_d$ are fundamentally linked, this directly translates to uncertainty in $C_d$. Thus, it is important to carefully consider the choice of these parameters when invoking ADBSat for such a calculation.

Currently, there are six possible options for the GSIM. Each outputs the drag, lift, pressure, and shear stress coefficients for a flat plate ($C_d$, $C_l$, $C_p$ and $C_{\tau}$ respectively). Only one pair of  \{${C_d,C_l}$\} and \{${C_p,C_\tau}$\} is calculated directly, with matrix \cref{eq:cdcl,eq:cpctau} being used to convert between the pairs using the angle between the oncoming flow and surface normal vector $\delta$.

\begin{gather}
 \begin{pmatrix} C_d & C_l \end{pmatrix}
 =
\begin{pmatrix} C_p & C_\tau \end{pmatrix}
  \begin{pmatrix}
   \cos(\delta) & \sin(\delta) \\
   \sin(\delta) & \cos(\delta) 
   \end{pmatrix}
\label{eq:cdcl}
\end{gather}

\begin{gather}
 \begin{pmatrix} C_p & C_\tau \end{pmatrix}
 =
\begin{pmatrix} C_d & C_l \end{pmatrix}
  \begin{pmatrix}
   \cos(\delta) & \sin(\delta) \\
   \sin(\delta) & -\cos(\delta) 
   \end{pmatrix}
\label{eq:cpctau}
\end{gather}

The six GSIMs available are:

\begin{itemize}




\item \textbf{Sentman:} A single accommodation coefficient is used to quantify energy transfer to the surface. Diffuse re-emission is assumed. A more realistic velocity distribution for incoming particles is used than that of more simplistic models, such as the Maxwell model \citep{DragModelling}. It also accounts for the relative motion of the surface and atmosphere. It is most accurate when the assumption of complete diffuse re-emission is correct, which is true for VLEO. Many literature sources exist that employ this model in analysing satellite flight, giving probable values for $\alpha_E$ at different atmospheric conditions \citep{MoeMoe2005, sentman, MoeMoe1998,AccommodationCoeffModel}. Thus, this model is widely used for the examination of satellite flight in VLEO. \Cref{eq:Cp_sent,eq:Ctau_sent} detail the calculation of $C_p$ and $C_\tau$ respectively, requiring $\alpha_E$, $\delta$, the speed ratio $s$, the wall temperature $T_w$, and the incident temperature $T_i$. The error function $\mathrm{erf}(x)$ is also used, and is described in \cref{eq:erf}.

\begin{multline} \label{eq:Cp_sent}
    C_p = \left( \cos^2(\delta) + \frac{1}{2s^2} \right) \bigg( 1 + \mathrm{erf}(s \cos(\delta)) \bigg) + \frac{\cos(\delta)}{s\sqrt{\pi}} e^{-s^2 \cos^2(\delta)} \\ + \frac{1}{2}\sqrt{\frac{2}{3} (1 + \frac{\alpha_E T_w}{T_i - 1})} \left[ \sqrt{\pi} \cos(\delta) \bigg( 1 + \mathrm{erf}(s \cos(\delta)) \bigg) + \frac{1}{s}  e^{-s^2 \cos^2(\delta)} \right]
\end{multline}

\begin{equation} \label{eq:Ctau_sent}
    C_\tau = \sin(\delta) \cos(\delta) \bigg( 1 + \mathrm{erf}(s \cos(\delta)) \bigg) + \frac{\sin(\delta)}{s \sqrt{\pi}} e^{-s^2 \cos^2(\delta)}
\end{equation}

\begin{equation} \label{eq:erf}
    \mathrm{erf}(x) = \frac{2}{\sqrt{\pi}} \int_{0}^{x} e^{-t^2} dt
\end{equation}

\item \textbf{Schaaf and Chambre:} Unlike the Sentman model which uses one accommodation coefficient, this model uses two: one each for tangential and normal momentum transfer. A more thorough description of the forces on the surface is thus possible \citep{DragModelling, SchaafCham}. However, it is often more difficult to obtain realistic values for the two accommodation coefficients than for Sentman's one. Therefore, if suitable accommodation coefficients are available, it is recommended to use this model. If not, Sentman should be employed instead. Finding $C_p$ and $C_\tau$ requires the normal and tangential momentum accommodation coefficients $\sigma_N$ and $\sigma_T$, the ambient temperature $T_{inf}$, $s$, $T_w$, and $\delta$. \Cref{eq:Cp_schaaf,eq:Ctau_schaaf} detail their calculation.

\begin{multline} \label{eq:Cp_schaaf}
    C_p = \frac{1}{s^2} \bigg[ \left(\frac{2-\sigma_N}{\sqrt{\pi}}  s \cos(\delta)  + \frac{\sigma_N}{2} \sqrt{\frac{T_w}{T_{inf}}} \right) e^{-s^2 \cos^2(\delta)} +  \\ \left( [2-\sigma_N]\left[  s^2 \cos^2(\delta) + \frac{1}{2} \right] + \frac{\sigma_N}{2} \sqrt{\frac{\pi T_w }{T_{inf}}} s \cos(\delta)  \right) \bigg( 1 + \mathrm{erf}(s \cos(\delta)) \bigg) \bigg]
\end{multline}

\begin{equation} \label{eq:Ctau_schaaf}
    C_\tau = \frac{\sigma_T \sin(\delta)}{s \sqrt{\pi}} \bigg[ e^{-s^2 \cos^2(\delta)} + s \sqrt{\pi} \cos(\delta) \bigg( 1 + \mathrm{erf}(s \cos(\delta)) \bigg) \bigg]
\end{equation}

\item \textbf{Cercignani-Lampis-Lord (CLL):} This model also requires two accommodation coefficients, one for tangential momentum accommodation and one for normal energy accommodation \cite{CLL_Lord}. It is primarily intended for DSMC applications, for which its mathematical formulation renders it particularly well suited. Its basis lies in a complex scattering kernel, for which closed-form solutions are not directly known. Instead, the closed-form solutions are based on modified expressions of the Schaaf and Chambre model that approximate the $C_d$ output of the model when it is applied in DSMC \cite{sabrina}. Therefore, as the ADBSat implementation relies on these approximated closed-form solutions, it is not recommended to use this model independently. Its use should be restricted to cases for which comparable DSMC simulations, also employing the CLL model, are available.

In this model, $C_p$ and $C_\tau$ are calculated for each individual species in the atmosphere, denoted by the subscript $j$. The species parameters $\beta_j$, $\gamma_j$, $\delta_j$ and $\zeta_j$ are are available in literature for each species \cite{Walker2014_CLL}. As well as the aforementioned accommodation coefficients $\sigma_T$ and $\alpha_N$, $T_w$ and $T_{inf}$ are also required.

\begin{multline} \label{eq:CLL_gamma1}
    \Gamma_1 = \frac{1}{\sqrt{\pi}} \bigg[ s \cos(\delta) e^{-s^2 \cos^2(\delta)} + \\ \frac{\sqrt{\pi}}{2} \bigg( 1 + 2 s^2 \cos^2(\delta) \bigg) \bigg( 1 + \mathrm{erf}(s \cos(\delta)) \bigg) \bigg]
\end{multline}

\begin{equation} \label{eq:CLL_gamma2}
    \Gamma_2 = \frac{1}{\sqrt{\pi}} \left[ e^{-s^2 \cos^2(\delta)} + s\sqrt{\pi} \cos(\delta) \bigg( 1 + \mathrm{erf}(s \cos(\delta)) \bigg) \right]
\end{equation}

If $\alpha_N < 1$: 

\begin{multline} \label{eq:Cp_CLL}
    C_{p,j} = \frac{1}{s^2} \Bigg[ \left( 1 + \sqrt{1 - \alpha_N} \right) \Gamma_1 + \\ \frac{1}{2} \left( e^{-\beta_j (1 - \alpha_N)^{\gamma_j}} \left( \frac{T_w}{T_{inf}} \right)^{\delta_j} \frac{\zeta_j}{s}   \right) \left( \sqrt{\frac{T_w}{T_{inf}}} \sqrt{\pi} \Gamma_2  \right) \Bigg]
\end{multline}

\begin{equation} \label{eq:Ctau_CLL}
    C_{\tau, j} = \frac{ \sigma_T \sin(\delta) }{s} \Gamma_2
\end{equation}

In other conditions (i.e. $\alpha_N = 1$), the equation for $C_{p,i}$ is the same as \cref{eq:Cp_schaaf}, with $\alpha_N$ substituted for $\sigma_N$. \Cref{eq:Ctau_CLL} remains unchanged. Finally, the total aerodynamic coefficients are computed as the weighted sum of the coefficients for each molecular species. An example of this for $C_p$ is shown in \cref{eq:CLL_sum}. $M_{avg}$ is the average mass of the mixture, $\chi_j$ is the species mole fraction, $m_j$ is the species mass, and $C_{p,j}$ is the species pressure coefficient. 

\begin{equation}\label{eq:CLL_sum}
    C_p = \frac{1}{M_{avg}} \sum_{j=1}^{N} \chi_j m_j C_{p,j}
\end{equation}


\item \textbf{Storch:} A key assumption of this model is 
hyperthermal flow, which exists when the ratio of flow velocity to molecular thermal velocity is small. Tangential and normal momentum accommodation are treated through separate accommodation coefficients, facilitating a more thorough description of the aerodynamic forces \cite{storchHyp}. However, while in some cases the hyperthermal assumption is valid in VLEO, care must be taken to ensure that the error introduced by neglecting particle thermal velocities is not significant \cite{sabrina}. Furthermore, as with the Schaaf and Chambre model, the two accommodation coefficients must be carefully chosen. Therefore, this model is not recommended for use unless the parameters of the case are well defined, and the user is confident in both the choice of accommodation coefficients and the hyperthermal conditions. $C_p$ and $C_\tau$ for this model are detailed in equations \cref{eq:Cp_storch,eq:Ctau_storch}, requiring $\delta$, $\sigma_N$, $\sigma_T$, the incident velocity $V$, and the average normal velocity of diffusely reflected molecules $V_w$. For all backward facing panels (i.e. $\delta > \frac{\pi}{2}$), both coefficients are set to zero.

\begin{equation} \label{eq:Cp_storch}
    C_p = 2 \cos(\delta) \left( \sigma_N \frac{V_w}{V} + [2-\sigma_N] \cos(\delta)  \right)
\end{equation}

\begin{equation} \label{eq:Ctau_storch}
    C_\tau = 2 \sigma_T \sin(\delta) \cos(\delta) 
\end{equation}

\item \textbf{Cook:} A basis in the Storch GSIM means that this model also deals only with hyperthermal flow. However, it introduces the simplification of only one accommodation coefficient. \citep{MostazaThesis, Cook}. As with the Storch model, it is important to ensure hyperthermal flow conditions are applicable before selecting this model for analysis. Furthermore, common errors must be avoided, such as the confusion of kinetic and atmospheric temperatures \cite{DragModelling,koppenwallner}. As the Sentman model converges to this model for hyperthermal conditions, the Sentman model exhibits a broader range of applications, and is recommended instead. The parameters $\delta$, $\alpha_E$, $T_w$ and $T_{inf}$ are required to calculate $C_d$ and $C_l$, as shown in \cref{eq:Cd_cook,eq:Cl_cook}.

\begin{equation} \label{eq:Cd_cook}
    C_d = 2 \cos(\delta) \left( 1 + \frac{2}{3} \cos(\delta) \sqrt{1 + \frac{\alpha_E T_w}{T_{inf} - 1}} \right)
\end{equation}

\begin{equation} \label{eq:Cl_cook}
    C_l = \frac{4}{3} \sin(\delta) \cos(\delta)  \sqrt{1 + \frac{\alpha_E T_w}{T_{inf} - 1}}
\end{equation}

\item \textbf{Newton:} The gas-surface interaction is determined by Newton's laws of motion. The particles are assumed to be hard spheres, losing all normal momentum on collision with the surface. This is a valid approximation in hypersonic flow at large Mach numbers and moderately small deflection angles. However, these conditions do not normally occur in VLEO, and it is not recommended to use this model for the usual satellite flight conditions. The equation for $C_p$ of a hard sphere, based only on $\delta$, can bee seen in \cref{eq:Cp_new}. $C_\tau$ is always zero.

\begin{equation} \label{eq:Cp_new}
    C_p = 2 \cos^2(\delta)
\end{equation}

\begin{equation} \label{eq:Ctau_new}
    C_{\tau} = 0
\end{equation}

\end{itemize}

Thorough explanations and comparisons of the different GSIMs for different uses are widely available in literature \cite{sabrina,MostazaThesis,DragModelling,ComparingDragCoeffs_GSIs,MoeMoe2005,MoeMoe2011,MaxwellVsCLL,MaxwellVsCLL2,AtmosphericDensities_satelliteDrag}. As the choice of GSIM is key to the accuracy of the analysis, it is recommended that users of ADBSat are confident in and can justify the model chosen for their analysis.

\section{Methodology}

An important part of computational data estimation is the validity of the input model. In other words, it is important to ensure that the geometric model of the spacecraft input into ADBSat is accurate. Previous programs have attempted to define a body through some form of in-house geometry definition, for example providing a set of shapes that can be arranged to define a satellite \cite{FreeMat}. ADBSat differs in that it takes as an input an existing file format common in CAD modelling, the Wavefront file format (with a ``.obj" extension).

A Wavefront file represents the surface mesh by means of vertices and face elements. Each vertex is defined by a set of (x,y,z) coordinates, and each face element by a combination of three vertices. By definition this results in triangular faces, which ADBSat requires. Each line of the Wavefront file defines either a vertex (where the line begins with ``v”) or a face (where the line begins with ``f"). By default, vertices are stored in a counter-clockwise order; therefore, the surface normals are implicitly defined. A particularly useful feature of this file format is the ability to specify multiple materials (where the line begins with ``usemtl"). The user can specify a different accommodation coefficient for each material, by providing as an input a list of accommodation coefficients in the same order as the materials specified in the Wavefront file, rather than a single value. ADBSat processes the model into a MATLAB formatted data file (with a ``.mat" extension) and explicitly calculates the outward surface normal, $n_p$, surface area, $A_p$, and barycentre $b_p$ for each element of the triangular mesh.

The Wavefront file format is standard for the representation of polygonal data in ASCII form. As a result, most current CAD software programs can export directly into it. It is thus accessible to the wider engineering community, among which general CAD knowledge is widespread. Additionally, the same model can be used for different analyses or as an input into different programs, without needing to be approximated or converted. However, it should be noted that as the meshing is independent of ADBSat, the user is responsible for the quality of the mesh. At present, ADBSat has no built-in mesh quality controls. Checks such as the removal of free-floating features, duplicate vertices, and non-manifold faces are left to the user. In particular, ADBSat cannot handle zero-area faces, and will output NaN values when these are encountered, leading to meaningless results.

\begin{figure}[hbt]
    \centering
    \includegraphics[page=1,scale=0.6]{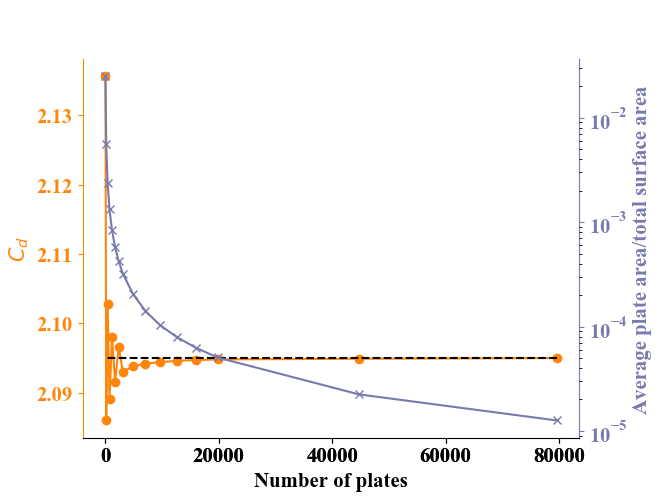}
    \caption{Analysis of the change in drag coefficient of a sphere as the number of plates used to represent it is increased. The closed-form Sentman solution is shown as a black dotted line, at $C_d = 2.095.$}
    \label{fig:nPlates}
\end{figure}

One important aspect to consider in creating a computational mesh is how accurately this mesh represents the desired shape, particularly for any rounded surfaces. The number of triangular plates it would take to represent such a surface exactly tends to infinity. However, the more plates used, the higher the computational time taken. Thus, there is a trade-off between computational time and accuracy of the model. Such a case can be seen in \cref{fig:nPlates} where a sphere of radius \SI{0.1}{\metre} using the Sentman model at \SI{200}{\kilo\metre} altitude is represented by 40 through 79600 triangular plates. It can clearly be seen that the result converges for an increasing number of plates to match that of the closed-form Sentman solution, while a lower number of plates introduces a high fluctuation to the results. It should be noted that even with 79600 plates, the runtime for each individual case is still quick, needing only a few seconds on a single core machine. Thus, as can be seen on the right-hand axis of \cref{fig:nPlates}, it is recommended for accuracy that rounded features have an average element area to total surface area ratio of \SI{1.5e-4} or lower. Below this point, the result is not particularly sensitive to fluctuations in the plate size. The consequences of the choice of plate size on the efficient of the shading algorithm is  discussed in a complementary work \cite{myValidationPaper}.

\begin{figure}[t]
    \centering
    \includegraphics[page=1,scale=0.3]{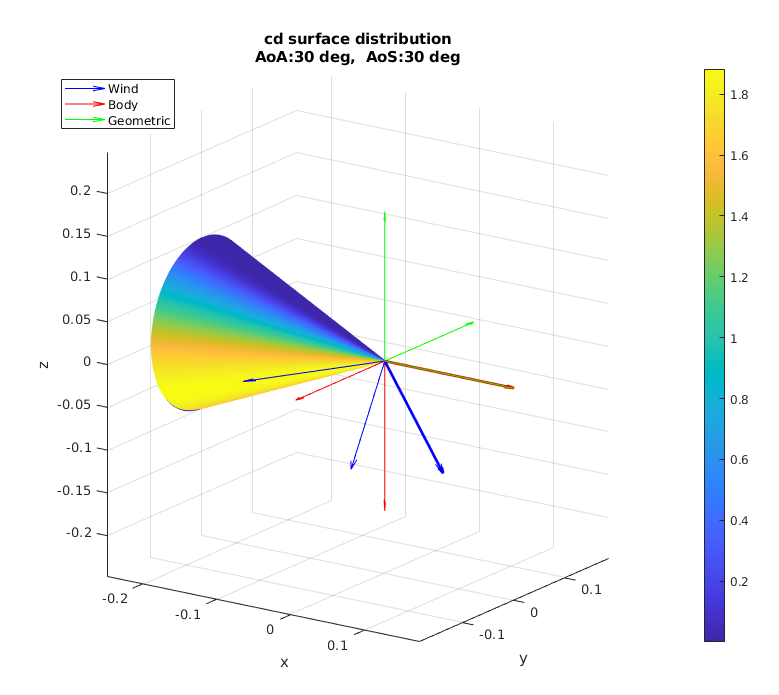}
    \caption{Geometric, body and wind frames for a cone at $\alpha$, $\beta =$ \SI{30}{\degree}. The x, y and z axes are aligned to the geometric reference frame.}
    \label{fig:coordFrames}
\end{figure}

In converting the model to its internal data file, ADBSat makes use of three reference frames: the geometric, body, and wind frames. Henceforth, the subscripts $g$, $b$ and $w$ respectively shall be used to refer to these frames. \Cref{fig:coordFrames} shows the three reference frames alongside a conical object pointing into the flow, with $\alpha$ and $\beta$ both equal to \SI{30}{\degree}. Both these parameters are taken as inputs from the user. Inputting a range for either or both of these will result in the calculation of an aerodynamic database.
 
 The geometric reference frame is that in which the Cartesian coordinates of the vertices are defined in the Wavefront file. By definition, the flow is aligned with the negative x-axis direction in the geometric frame when $\alpha$ and $\beta$ are both zero. These angles define the orientation of the body with respect to the flow.

 As is typical for flying bodies, the body frame is a right-handed co-ordinate frame with the positive direction of the z axis pointing towards the Earth. The transformation matrix used by ADBSat to define the body frame is $L_{gb}$, which transforms a vector expressed in the body frame into the geometric frame. It can be seen in \cref{eq:Lgb}. 

\begin{equation}
L_{gb} = \begin{pmatrix} 1 & 0 & 0\\ 0 & -1 & 0\\ 0 & 0 & -1 \end{pmatrix}
\label{eq:Lgb}
\end{equation}




For a given $\alpha$ and $\beta$, the matrix $L_{bw}$ transforms a vector from the wind frame into the body frame, as shown in \cref{eq:Lbw}.

\begin{equation}\label{eq:Lbw}
L_{bw} = \begin{pmatrix} \cos \alpha \cos \beta & - \cos \alpha \sin \beta &  - \sin \alpha \\ \sin \beta & \cos \beta & 0\\ \sin \alpha \cos \beta & - \sin \alpha \sin \beta & \cos \alpha \end{pmatrix}
\end{equation}

Thus, the transformation from the wind to the geometric frame will be given by $L_{gw}$ = $L_{gb}L_{bw}$. By using the definition of the wind frame, which is that the flow in this frame is always in the negative x-axis direction regardless of $\alpha$ and $\beta$, the velocity direction in the geometric frame, $\vec{n}_v$, can be calculated from \cref{eq:nv}. \Cref{eq:delta} shows how ADBSat calculates the angle $\delta_i$ between the velocity vector $\vec{n}_v$ and the surface normal of any element of the mesh $\vec{n}_{i}$. 

\begin{equation}\label{eq:nv}
    \vec{n}_v = L_{gw} \begin{bmatrix} -1 \\ 0 \\ 0 \end{bmatrix}
\end{equation}

\begin{equation}\label{eq:delta}
    \delta_i = \arccos(-\vec{n}_v \cdot \vec{n}_{i})
\end{equation}


By means of the chosen GSIM, this angle can be transformed into local pressure and shear stress coefficients for each panel, $c_{p,i}$ and $c_{\tau,i}$ respectively. These coefficients depend not only on the incidence angle, but also on a range of further parameters that characterise the gas-surface interaction and vary for different models. Finally, global force and moment coefficients for the body in geometric axes are obtained through \cref{eq:Cgf,eq:Cgm}.

\begin{equation} \label{eq:Cgf}
    \vec{C}^g_{fw} = \begin{bmatrix} C_{fwx} \\ C_{fwy} \\ C_{fwz} \end{bmatrix} = \frac{1}{A_{ref}}  \sum_{i=1}^{n} (c_{\tau,i} \vec{\tau}_i - c_{p,i} \vec{n}_{i})A_{i}
\end{equation}

\begin{equation} \label{eq:Cgm}
    \vec{C}^g_{mw} = \begin{bmatrix} C_{mwx} \\ C_{mwy} \\ C_{mwz} \end{bmatrix} = \frac{1}{A_{ref}L_{ref}} \sum_{i=1}^{n} \vec{r}_{i} \times (c_{\tau,i} \vec{\tau}_i - c_{p,i} \vec{n}_{i}) A_{i}
\end{equation}

In this equation, $A_{ref}$ is the reference surface area, the default being half of the mesh surface area. $L_{ref}$ is the reference length, defined as half the distance between the maximum and minimum x coordinates. $A_i$ is the area of each plate. Vector $\vec{r}_{i}$ points from the geometric moment reference point to the barycentre of triangular plate $i$. Vector $\vec{\tau}_i$ is a unit vector in the direction of shear stress, calculated from \cref{eq:tau_i}.

\begin{equation} \label{eq:tau_i}
\vec{\tau}_i = \vec{n}_{i} \times (\vec{n}_v \times \vec{n}_{i})
\end{equation}

It is left to the user to reference $\vec{C}_{fw}^g$ and $\vec{C}_{mw}^g$ to the relevant area and length for the problem, according to \cref{eq:Cgf2,eq:Cgm2}.

\begin{equation} \label{eq:Cgf2}
    \vec{C}_{fw2}^g = \frac{A_{ref}}{A_{ref2}} \vec{C}_{fw}^g
\end{equation}

\begin{equation} \label{eq:Cgm2}
    \vec{C}_{mw2}^{g} = \frac{A_{ref}L_{ref}}{A_{ref2}L_{ref2}} \vec{C}_{mw}^g
\end{equation}

The moment and force coefficients can be translated into body or wind frames using the transformation matrices in \cref{eq:Lgb,eq:Lbw}. In addition, moment coefficients are obtained using the origin of the the geometric frame as the moment reference centre (MRC), which does not generally coincide with the centre of gravity (CoG). \Cref{eq:COG} translates these coefficients to the CoG.

\begin{equation} \label{eq:COG}
    \vec{C}_{mw}^{CoG} = -\vec{r}_{CoG} \times \vec{C}_{fw} + \vec{C}_{mw}^{MRC}
\end{equation}

Finally, the vehicle moment and force coefficients in the body axes, $C_M^b$ and $C_F^b$ respectively, can be obtained using the transformation matrix defined in \cref{eq:Lgb}. This is explicitly shown in \cref{eq:CmLgb,eq:CfLgb}.

\begin{equation} \label{eq:CmLgb}
    C_M^b = L_{gb}^{-1} C_{mw}^g
\end{equation}

\begin{equation} \label{eq:CfLgb}
    C_F^b = L_{gb}^{-1} C_{fw}^g
\end{equation}

For clarity, these coefficients are defined in \cref{eq:CMb,eq:CFb}. Here, $C_l$, $C_m$ and $C_n$ are the roll, pitch and yaw aerodynamic moment coefficients respectively and $C_A$, $C_Y$ and $C_N$ are the axial, lateral and normal force coefficients respectively.

\begin{equation} \label{eq:CMb}
    C_M^b = \{ C_l, C_m, C_n \}
\end{equation}

\begin{equation} \label{eq:CFb}
    C_F^b = \{ C_A, C_Y, C_N \}
\end{equation}

Choosing the GSIM to be applied, in essence, determines which pressure and shear stress coefficient calculations to apply. Should the user wish to define a new model, the template of the existing model scripts can be followed: code should be written which calculates any parameters needed to ultimately calculate the force coefficients for the body according to the new model. Thus, the intended versatility of ADBSat is apparent, and the easy implementation of future breakthroughs in FMF modelling is assured.

\section{Shading Algorithm}

Unassisted, GSIMs cannot account for panels which are shielded from the flow by upstream features. This is often a source of error when calculating aerodynamic coefficients \cite{FreeMat}. ADBSat addresses this shortcoming by implementing a simple shadow analysis algorithm \cite{MostazaThesis}. A flowchart of the algorithm is shown in \cref{fig:shading_analysis_flowchart}. Fundamentally, it works as follows:

\begin{figure}[!t]
\centering
\includegraphics[width=0.7\linewidth]{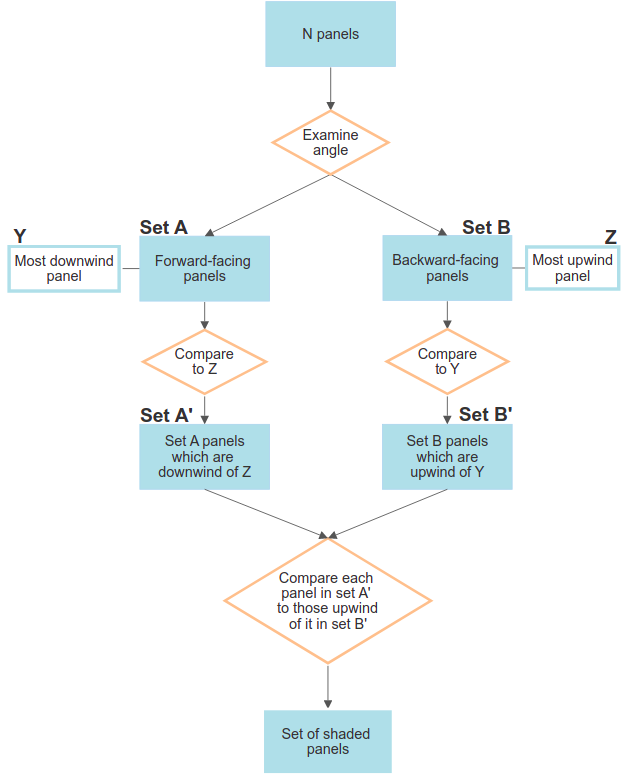}
\caption{A flowchart of the shading analysis algorithm.}
\label{fig:shading_analysis_flowchart}
\end{figure}

\begin{figure}[!t]
\centering
\includegraphics[width=0.7\linewidth]{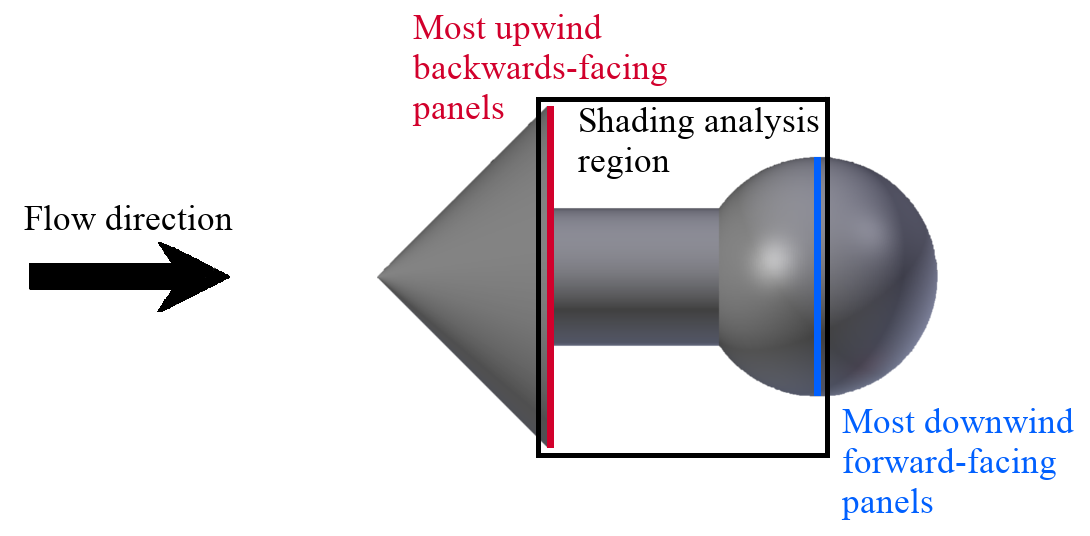}
\caption{An example of the shading analysis region of a test case.}
\label{fig:shading_analysis_explanation}
\end{figure}

\begin{enumerate}

    \item Panels are split into two sets based on the angle between their normal and the oncoming flow vector: 
    
    \begin{itemize}
        \item set A contains forward-facing panels and those parallel to the flow ($\delta \leq \frac{\pi}{2}$).
        \item set B contains backward-facing panels ($\delta > \frac{\pi}{2}$).
    \end{itemize}
    
    Only set A can be shadowed, and only set B can shadow other panels.
    
    \item The most downwind panel in set A is determined (panel Y). Similarly, the most upwind panel in set B is determined (panel Z).
    
    \item Only panels in set A that are downwind of Z can be shadowed. Similarly, only panels in set B upwind of Y can shadow other panels. Selecting only these panels reduces set A to set A', and set B to set B'. A pictorial representation of this is shown in \cref{fig:shading_analysis_explanation}.
    
    \item For each panel in set A', a sub-set of B' that are upwind of it is identified. The barycentre of the considered panel is then checked against a 2D projection of each panel in the sub-set. If its barycentre falls inside any of the projections, then the panel is marked as shadowed and its contributions to any aerodynamic coefficients are assigned a value of zero.
    
\end{enumerate}

This algorithm is an algebraic approximation based on geometric projections and is intended as a fast pseudo ray-tracer. In other words, it is not intended as a foolproof method of shading determination, but rather, a fast one.  As panels cannot be part-shaded, a coarse discretisation of the shape may lead to incorrect results. Shapes with large flat sides, for example solar arrays, are particularly susceptible to this. The extent of this effect will be examined and thoroughly discussed in future work.

\section{Example case: SOAR}

\begin{figure}[t]
     \centering
     \begin{subfigure}[b]{0.4\textwidth}
         \centering
         \includegraphics[width=\textwidth]{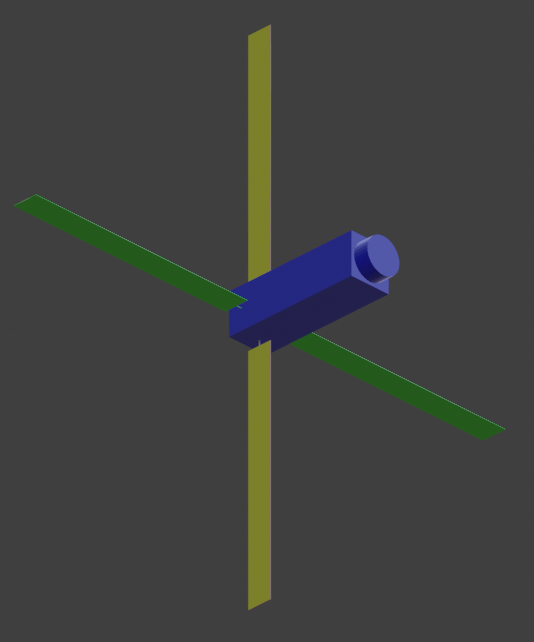}
         \caption{}
         \label{fig:SOAR1}
     \end{subfigure}
     \begin{subfigure}[b]{0.4\textwidth}
         \centering
         \includegraphics[width=\textwidth]{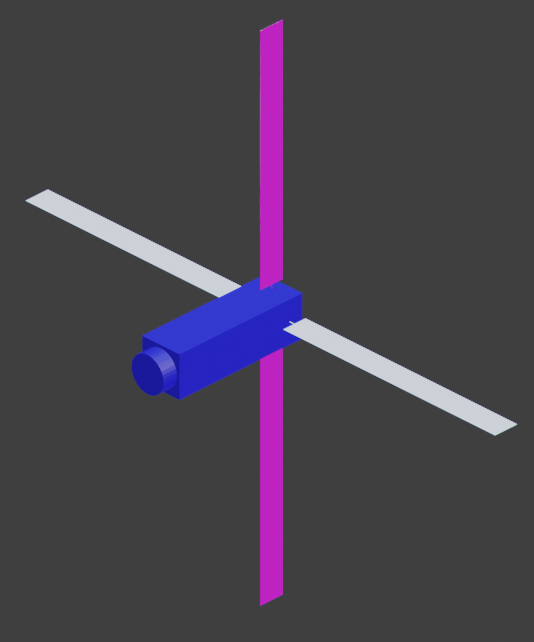}
         \caption{}
         \label{fig:SOAR2}
     \end{subfigure}
        \caption{The SOAR geometry viewed from two different angles. The different colours represent the five materials.}
        \label{fig:SOAR}
\end{figure}

\begin{figure}[t]
\centering
\includegraphics[width=0.6\linewidth]{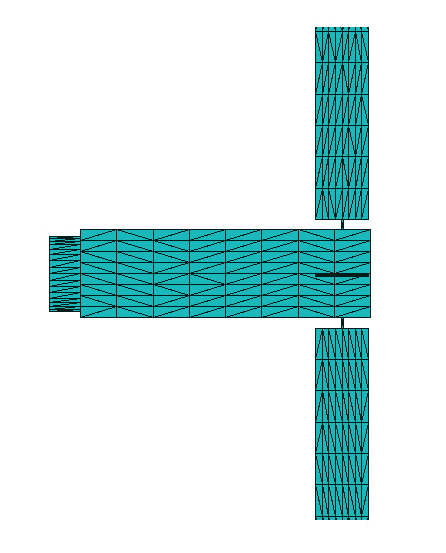}
\caption{A side-on close-up of the main body geometry, showing the individual triangular panels which make up the geometry. Only part of the steerable fins is visible.}
\label{fig:SOARpanels}
\end{figure}

The Satellite for Orbital Aerodynamic Research (SOAR) is a 3U CubeSat mission proposed to study the effects of different materials in VLEO \cite{SOAR}. It aims to do this by using four steerable fins that can expose four different materials to the flow in order to discern their aerodynamic properties. It also carries an ion and neutral mass spectrometer (INMS) that can determine the composition of the VLEO atmosphere in-situ. A simplified version of its proposed configuration, with the fins parallel to the body of the spacecraft, was modelled using the SolidWorks and Blender CAD programs. It is comprised of 3233 flat triangular plates, with the different materials - four test materials and one body material - labelled from 1 to 5. Because different CAD programs can be used to easily generate the geometries, the process of model generation should be relatively simple for most users. Importing this model into ADBSat takes less than one second. It can be seen in \cref{fig:SOAR}, from two different angles to show all five different-coloured materials. \cref{fig:SOARpanels} shows a side-on close-up of the body and part of the steerable fins after import into MATLAB. The individual triangular plates that make up the model can be seen therein.

The example case was examined at an altitude of \SI{200}{\kilo\metre}, using Sentman's model. Intermediate solar activity conditions were chosen, using a reference date of 19 January 2015 at 00:00:00 and latitude and longitude (0,0). Representative solar indices corresponding to these conditions were chosen as 81-day average $F_{10.7} = 138.1$ and daily $F_{10.7} = 121.7$. $A_p$ magnetic indices were in the range \SIrange{2.9}{9.0}{}. A full aerodynamic and solar database was obtained for the object, with incidence angles in the range $-90 < \alpha < 90$ and $-180 < \beta < 180$, with a step size of one degree, resulting in 65341 individual combinations. Shading analysis was enabled. The total time needed to produce this database without graphical output was around 1.25 hours on an Intel\textsuperscript{\tiny\textregistered} Core\texttrademark \ i7 vPro\textsuperscript{\tiny\textregistered} quad-core machine. The mean time to run each calculation loop was approximately \SI{0.07}{\second}.

The aerodynamic database is stored in the aforementioned MATLAB workspace file. Here, the aerodynamic and solar force and moment coefficients for each combination of $\alpha$ and $\beta$ can be accessed. As previously mentioned, for all calculations the reference area is half of the total surface area, and the reference length is half of the object size along the X axis. The projected area at each combination of angles is also stored in the output file. The value of drag coefficient can then be obtained by referencing the aerodynamic force coefficient along the X axis in the wind co-ordinate system to the projected area, using \cref{eq:Cgf2}.

\begin{figure}[t]
\centering
\includegraphics[width=0.6\linewidth]{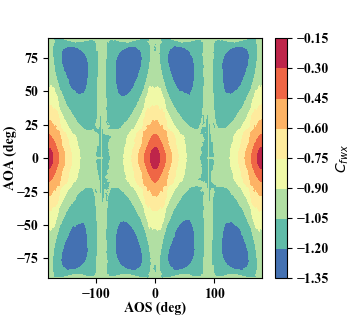}
\caption{Aerodynamic coefficient along the x-axis for the example case where all surfaces have $\alpha_E$ = 1. The reference area is half of the total surface area of the body.}
\label{fig:example_aero_alpha1}
\end{figure}

For the example case, two aerodynamic databases have been calculated. The first uses $\alpha_E$ = 1 for all five materials. As the value of $\alpha_E$ is the only material characteristic taken into account, functionally this simplifies the model to having the same material across the entire surface area. A contour plot of the resulting force coefficient in the X axis direction of the wind frame can be seen in \cref{fig:example_aero_alpha1}. The values used to produce the plot are unchanged from the ADBSat outputs. It can be seen that as the value of $\alpha_E$ does not vary across the different materials, the plot is symmetric.

\begin{figure}[t]
\centering
\includegraphics[width=0.6\linewidth]{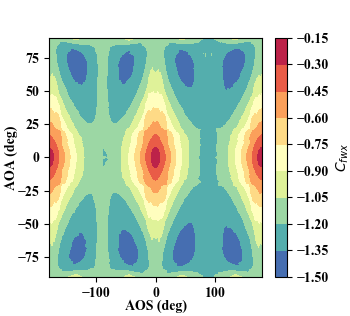}
\caption{Aerodynamic coefficient along the x-axis for the example case where surfaces have varying values of $\alpha_E$. The reference area is half of the total surface area of the body.}
\label{fig:example_aero_alphadiff}
\end{figure}

The second case is used to demonstrate variation in the gas-surface interaction properties, representing the lack of knowledge of the materials present on the flight model of SOAR. A value of $\alpha_E$ = 1 is used for the body of the CubeSat and the four materials on the fins are given values of $\alpha_E \in [0.95, 0.9, 0.85, 0.8]$. As can be seen in \cref{fig:example_aero_alphadiff}, this results in a more negative aerodynamic force coefficient, which can also be interpreted as higher drag. An asymmetric distribution with angle of attack and sideslip is also observed due to the difference in $\alpha_E$ between the different fins. As aerodynamic force coefficients are particularly susceptible to changes in $\alpha_E$, this is expected.

Evaluating the accuracy of this output, particularly with regard to the drag characteristics of the object, is an important part of establishing the usability of ADBSat. Verification and validation of the software was undertaken through comparison with established methods of drag analysis and published aerodynamic coefficient values. The final set of test shapes is large, comprising of 14 test shapes analysed across a range of atmospheric conditions and mesh configurations. The performance of the novel shading algorithm is thoroughly tested, as well as the influence of multiple reflections on the results. Five past satellite missions with publicly available aerodynamic data are also used to verify the program outputs. This extensive validation is thoroughly detailed in a separate paper which focuses on an explanation of our cases, methods, and results \cite{myValidationPaper}.

\section{Conclusions}
\label{sec:Conclusions}

Much like other similar programs, ADBSat is designed to provide a fast, accurate approximation of the aerodynamic properties of spacecraft in free-molecular flow. It improves on previous implementations by exploiting existing knowledge of CAD modelling in the aerodynamic engineering community, taking as an input a model of the spacecraft which can be made in most common CAD software suites. This reduces the learning necessary to use the software, as well as the time needed for model design. It also implements a novel shading algorithm based on 2-dimensional projections of the triangular plates that make up the model. Additionally, the flat-plate representation of the spacecraft has been harnessed to calculate the solar coefficients, if required. Once validated, this will extend the utility of the software beyond the orbital regime in which aerodynamic forces dominate and into that in which solar radiation pressure dominates. Its modular design also means that new GSIMs and solar coefficient models can be easily implemented in the future.

The validation of the SRP model, and addition of further GSIM and SRP models, is undoubtedly the most significant possible future improvement to the program. This will extend the range of orbital cases for which the program can be applied, both in terms of the aerodynamics and the SRP conditions, and allow it to remain up-to-date with further developments in the field. The addition of mesh compatibility and quality checks will also be explored in the future, to aid users in obtaining useful outputs from the software. Furthermore, auxiliary functionality can be added to aid the user in modelling their case - for example, the program currently does not have the capability to estimate accommodation coefficients, which need to be input by the user. Integrating the Langmuir isotherm model \cite{AccommodationCoeffModel} would provide an estimate based on the input atmospheric conditions. While this model is not universally applicable, some users may find it useful, and those for whom it is not could simply bypass it and input their own value(s). 

The main advantage of ADBSat is the speed with which it outputs results, with a full aerodynamic and solar database for a model comprising of over 3000 elements taking around 1.25 hours to output. Although the runtime increases with the number of panels, this is still far more time-efficient than other existing methods. Thus, it can complement current existing methods such as direct simulation Monte Carlo (DSMC), particularly for preliminary design cases where there is a need for large aerodynamic databases for which this method is time-consuming. It can also be employed in cases in which there are many feasible satellite geometries being considered, such as at the early mission design stage. Additionally, it can be used for post-mission analysis and data interpretation. To summarise, this program provides a fast, practical solution for the aerodynamic analysis of satellite bodies, with a flexible approach which is conducive to the easy implementation of future advancements in atmospheric physics models.

\section*{Acknowledgements}

L. Sinpetru would like to thank her supervisors and colleagues for their valuable discussion. This work was supported by doctoral funding provided by the University of Manchester.

\section*{CRediT authorship statement}
\noindent \textbf{Luciana Sinpetru:} Software, Writing - original draft, Writing - Review \& Editing, Visualization 

\noindent \textbf{Nicholas H. Crisp:} Software, Writing - Review \& Editing 

\noindent \textbf{David Mostaza-Prieto:} Conceptualization, Methodology, Software 

\noindent \textbf{Sabrina Livadiotti:} Software 

\noindent \textbf{Peter C. E. Roberts:} Supervision, Funding Acquisition

\bibliographystyle{elsarticle-num-names}
\bibliography{sample.bib}

\end{document}